\documentclass[12pt]{article}
 \usepackage{graphicx,amsmath,amssymb}
\usepackage{indentfirst}
 \usepackage[usenames]{color}
 \usepackage[colorlinks=true, urlcolor=navyblue, linkcolor=navyblue, citecolor=navyblue]{hyperref}
\usepackage{epstopdf}
\usepackage{appendix}

\definecolor{navyblue}{rgb}{0,0.08,0.45}

\def\Dslash{\raise.15ex\hbox{/}\kern-.7em D}
\def\Pslash{\raise.15ex\hbox{/}\kern-.7em P}

\newcommand{\beq}{\begin{equation}}
\newcommand{\enq}{\end{equation}}
\newcommand{\beqa}{\begin{eqnarray}}
\newcommand{\beqast}{\begin{eqnarray*}}
\newcommand{\enqa}{\end{eqnarray}}
\newcommand{\enqast}{\end{eqnarray*}}
\newcommand{\beml}{\begin{multline}}
\newcommand{\enml}{\end{multline}}

\newcommand{\bec}{\begin{center}}
\newcommand{\enc}{\end{center}}
\newcommand{\beqo}{\begin{quote}}
\newcommand{\enqo}{\end{quote}}

\newcommand{\half}{{\textstyle{\frac{1}{2}}}}

\newcommand{\mbf}[1]{\mathbf{#1}}

\begin{document}

\begin{flushright}
{
\small
SLAC-PUB-15715\\
\date{today}}
\end{flushright}

\vspace{60pt}

\centerline{\Large \bf Light-Front Holographic QCD and the Confinement Potential}

\vspace{20pt}


\centerline{{
Stanley J. Brodsky,$^{a}$ 
\footnote{E-mail: \href{mailto:sjbth@slac.stanford.edu}{sjbth@slac.stanford.edu}}
Guy F. de T\'eramond,$^{b}$ 
\footnote{E-mail: \href{mailto:gdt@asterix.crnet.cr}{gdt@asterix.crnet.cr}}
and
Hans G\"unter Dosch$^{c}$ 
\footnote{E-mail: \href{mailto:gdt@asterix.crnet.cr}{dosch@thphys.uni-heidelberg.de}}
}}

\vspace{20pt}

{\centerline {$^{a}${\it SLAC National Accelerator Laboratory, 
Stanford University, Stanford, CA 94309, USA}}

\vspace{4pt}

{\centerline {$^{b}${\it Universidad de Costa Rica, San Jos\'e, Costa Rica}}

\vspace{4pt}

{\centerline {$^{c}${\it Institut f\"ur Theoretische Physik, Philosophenweg 16, D-6900 Heidelberg, Germany}}

 \vspace{20pt}
 
 \begin{abstract}

Light-Front Hamiltonian theory, derived from the quantization of the QCD Lagrangian at fixed light-front time $\tau = t+z/c$, provides a rigorous frame-independent framework for solving nonperturbative QCD.  The eigenvalues of the light-front QCD Hamiltonian predict the hadronic mass spectrum, and the corresponding eigensolutions provide the light-front wavefunctions which describe hadron structure. The valence Fock-state wavefunctions of the light-front QCD Hamiltonian satisfy a single-variable relativistic equation of motion, analogous to the nonrelativistic radial Schr\"odinger equation, with an effective confining potential $U$ which systematically incorporates the effects of higher quark and gluon Fock states.  In fact, the potential $U$ has a unique form if one requires that the action for zero quark mass  remains conformally invariant. We also show that the holographic mapping of gravity in AdS space to QCD with a specific soft-wall dilaton yields the same light-front Schr\"odinger equation.    Light-front holography also leads to a precise relation between the bound-state amplitudes in the fifth dimension $z$ of AdS space and the boost-invariant light-front wavefunctions describing the internal structure of hadrons in physical space-time. The elastic and transition form factors of the pion and the nucleons are found to be well described in this framework. The predictions of the LF equations of motion include a zero-mass pion in the chiral $m_q\to 0$ limit, and linear Regge trajectories ${M}^2(n,L) \propto n+L$  with the same slope  in the
radial quantum number $n$ and orbital angular momentum $L$. The light-front AdS/QCD holographic approach thus gives a frame-independent representation of color-confining dynamics, Regge spectroscopy, and the excitation spectra of relativistic light-quark meson and baryon bound states in QCD in terms of a single mass parameter.
We also briefly discuss the implications of the underlying conformal template of QCD  for renormalization scale-setting and the implications of light-front quantization for the value of the cosmological constant.

\end{abstract}




\section{Introduction}

If one sets the masses of the quarks to zero, no mass scale appears explicitly in the QCD Lagrangian.  The classical theory thus displays invariance under both scale  (dilatation) and special conformal transformations~\cite{Parisi:1972zy}.   Nevertheless, the quantum theory built upon this conformal template displays color confinement,  a mass gap, as well as asymptotic freedom. A  fundamental question is how does the mass scale which determines the masses of the light-quark hadrons,  the range of color confinement, and running of the  coupling appears in QCD?

A hint to the origin of the mass scale in nominally conformal theories was given in 1976 in a remarkable paper by  V.~de Alfaro, S.~Fubini and G.~Furlan (dAFF)~\cite{deAlfaro:1976je} in the context of one-dimensional quantum mechanics.  They showed that the mass scale which 
breaks the dilatation invariance can appear in the equations of motion, while retaining  the conformal invariance of the action.  In fact, this is only possible if the resulting potential has the form of a confining harmonic oscillator and the transformed time variable $\tau$ that appears in the confining theory has a limited range.

In these proceedings we will show that the dAFF procedure,
together with light-front quantum mechanics and light-front holographic mapping,
leads to a remarkable analytic approximation to QCD -- a light-front Hamiltonian and corresponding one-dimensional light-front (LF) Schr\"odinger and Dirac equations which are frame-independent, relativistic, and reproduce crucial features of the spectroscopy and dynamics of the light-quark hadrons.
The predictions of the LF equations of motion include a zero-mass pion in the chiral $m_q\to 0$ limit, and linear Regge trajectories ${M}^2(n,L) \propto n+L$  with the same slope  in the
radial quantum number $n$ (the number of nodes) and 
$L= {\rm max} \,|L^z|$, the internal orbital angular momentum.    We will also show that the effective  confinement potential which appears in the LF equations of motion is unique.
 
The quantization of QCD at fixed light-front  time $\tau= t +z/c$ ~\cite{Dirac:1949cp} (Dirac's Front Form)  provides a first-principles Hamiltonian method for solving nonperturbative QCD. 
It is rigorous, has no fermion-doubling, is formulated in Minkowski space, and it is frame-independent.  
Solving nonperturbative QCD is equivalent to solving the light-front Heisenberg matrix eigenvalue problem.   Angular momentum $J^z$ is conserved at every vertex.
Given the boost-invariant light-front wavefunctions $\psi_{n/H}$ (LFWFs), one can
compute a large range of hadron
observables, starting with structure functions, generalized parton distributions, and form factors, as well as  providing a  quantum-mechanical probabilistic interpretation of hadronic states. 
It is also possible to compute jet hadronization at the amplitude level from first principles from the LFWFs~\cite{Brodsky:2008tk}. The LFWFs of hadrons thus provide a direct connection between observables and the QCD Lagrangian. 

The simplicity of the front form contrasts with the usual instant-form formalism. Current matrix elements defined at ordinary time $t$ must include the coupling of photons and vector bosons fields  to connected vacuum-induced currents; otherwise, the result is not Lorentz-invariant.  Thus the knowledge of the hadronic eigensolutions of the instant-form Hamiltonian are insufficient for determining form factors or other observables.   In addition, the boost of an instant form wavefunction from $p$ to $p+q$ changes particle number and is an extraordinarily complicated dynamical problem. 

Light-front quantization thus provides an ideal framework to describe the internal constituent structure of hadrons since the formalism is rigorous, relativistic and frame-independent.   The light-front wavefunctions derived from the hadronic eigensolutions are the natural relativistic extension of the familiar non-relativistic  Schr\"odinger wavefunctions of atomic physics. 

\section{The Light-Front Vacuum}

It is conventional to define the vacuum in quantum field theory as the lowest energy eigenstate of the instant-form Hamiltonian.  Such an eigenstate is defined at a single time $t$ over all space $\vec x$.  It is thus acausal and frame-dependent. The instant-form vacuum thus must be normal-ordered in order to avoid violations of causality when computing correlators and other matrix elements.

In contrast, in the front form, the vacuum state is defined as the  eigenstate of lowest invariant mass $M$. It is defined at fixed light-front time $x^+ = x^0 + x^3$ over all $x^-= x^0 - x^3$ and $\vec x_\perp$, the extent of space that can be observed within the speed of light.   It is frame-independent and only requires information within the causal horizon. The LF vacuum for QED, QCD, and even the Higgs Standard Model is trivial up to zero modes. In this sense it is already normal-ordered.   
 
There are thus no quark or gluon vacuum condensates in the LF vacuum-- as first noted by Casher and Susskind~\cite{Casher:1974xd}; the corresponding physics is contained within the 
LFWFs themselves~\cite{Brodsky:2009zd,Brodsky:2010xf,Chang:2013pq}, thus eliminating a major contribution to the cosmological constant.   
In the light-front,
phenomena such as the GMOR relation -- usually  associated with condensates in the instant form vacuum --  are properties of the the hadronic LF wavefunctions themselves.  
In the case of the Higgs theory, the usual Higgs VEV is replaced by a classical $k^+=0$  background zero-mode field 
which is not sensed by the energy momentum tensor~\cite{Srivastava:2002mw}.  The front-form vacuum is thus a match to the ``void" -- the observed universe without matter.   
Thus it is natural in the front form to obtain zero cosmological constant.

\section{The Conformal Symmetry Template}

In the case of perturbative QCD, the running coupling  $\alpha_s(Q^2)$ becomes constant in the limit of zero $\beta$-function and zero quark mass, and conformal symmetry becomes manifest.  In fact, the renormalization scale uncertainty in 
pQCD predictions can be eliminated by using the Principle of Maximum Conformality (PMC)~\cite{Brodsky:2011ig}.
Using the PMC/BLM procedure~\cite{Brodsky:1982gc}, 
all non-conformal contributions in the perturbative expansion series are summed into the running coupling by shifting the renormalization scale in $\alpha_s$ from its initial value, and one obtains unique, scale-fixed, scheme-independent predictions at any finite order.  One can also introduce a generalization of conventional dimensional regularization which illuminates the renormalization scheme and scale ambiguities of pQCD predictions, exposes the general pattern of nonconformal terms, and allows one to systematically determine the argument of the running coupling order by order in pQCD in a form which can be readily automatized~\cite{Mojaza:2012mf, Brodsky:2013vpa}
The resulting PMC scales and  finite-order PMC predictions are both to high accuracy independent of the choice of initial renormalization scale.  For example, PMC scale-setting leads to a scheme-independent 
pQCD prediction~\cite{Brodsky:2012ik}
for the top-quark forward-backward asymmetry which is within one $\sigma$ of the Tevatron measurements. 
The PMC procedure also provides scale-fixed, scheme-independent commensurate scale relations~\cite{Brodsky:1994eh}, relations between observables which are based on the underlying conformal behavior of QCD such as the generalized Crewther relation~\cite{Brodsky:1995tb}.
The PMC satisfies  all of the principles of the renormalization group: reflectivity, symmetry, and transitivity, and it thus eliminates an unnecessary source of systematic error in pQCD predictions~\cite{Wu:2013ei}.

\section{Light-Front Holography}

An important  analysis tool for QCD is Anti-de Sitter space in five dimensions.  In particular, AdS$_5$ provides a remarkable geometric representation of the conformal group which 
underlies the conformal symmetry of classical QCD. One can modify AdS space by using a dilaton factor in the AdS
action 
$e^{\varphi(z)} $ to introduce the QCD 
confinement scale.  However, we shall show that
if one imposes the requirement that  the action of the corresponding one-dimensional effective theory  remains conformal invariant,
 this is only possible if  the dilaton profile  $\varphi(z) \propto z^s$ is constrained to have the specific power $s = 2$,  a remarkable result which follows from the dAFF construction of conformally invariant quantum mechanics~\cite{Brodsky:2013kpr}.     The quadratic form 
$\varphi(z) = \pm \, \kappa^2 z^2$
immediately leads to linear Regge trajectories~\cite{Karch:2006pv} in the hadron mass squared.   

Light-Front Holography refers to the remarkable fact that dynamics in AdS space in five dimensions is dual to 
a semiclassical approximation to
Hamiltonian theory in physical  $3+1$ space-time quantized at fixed light-front time~\cite{deTeramond:2008ht}.  The  correspondence between AdS and QCD, which was originally motivated by
the AdS/CFT correspondence between gravity on a higher-dimensional 
space and conformal field theories 
in physical space-time~\cite{Maldacena:1997re},  has its most explicit and simplest realization as a direct holographic mapping to light-front Hamiltonian theory~\cite{deTeramond:2008ht}. 
For example, the equation of motion for mesons on the light-front has exactly the same single-variable form as the AdS/QCD equation of motion; one can then interpret the AdS fifth dimension variable $z$ in terms of the physical variable $\zeta$, representing the invariant separation of the $q$ and $\bar q$ at fixed light-front time.  There is a precise 
connection between the quantities that enter the fifth 
dimensional AdS space and the physical variables of LF theory.  The AdS mass  parameter $\mu R$ maps to the LF orbital angular momentum.  The formulae for electromagnetic~\cite{Polchinski:2002jw} and gravitational~\cite{Abidin:2008ku} form factors in AdS space map to the exact Drell-Yan-West formulae in light-front QCD~\cite{Brodsky:2006uqa, Brodsky:2007hb, Brodsky:2008pf}.  

As we shall discuss, the light-front holographic principle provides a precise relation between the bound-state amplitudes in AdS space and the boost-invariant LF wavefunctions describing the internal structure of hadrons in physical space-time. See Fig. \ref {dictionary}. The resulting valence Fock-state wavefunctions
satisfy a single-variable relativistic equation of motion analogous to the eigensolutions of the nonrelativistic radial Schr\"odinger equation. 
The quadratic dependence in the effective quark-antiquark potential 
$U(\zeta^2,J) =  \kappa^4 \zeta^2 +2 \kappa^2(J-1)$ 
is determined uniquely from conformal invariance.   The  constant term $ 2 \kappa^2(J-1) = 2 \kappa^2(S+L-1)  $  is fixed by the duality between AdS and LF quantization for spin-$J$ states, a correspondence which follows specifically from the separation of kinematics and dynamics on the light-front~\cite{deTeramond:2013it}.  
The LF potential thus has a specific power dependence--in effect, it is a light-front harmonic oscillator potential.
It  is confining and reproduces the observed linear Regge behavior of the light-quark hadron spectrum in both the orbital angular momentum $L$ and the radial node number $n$. The  pion is predicted to be massless in the chiral limit ~\cite{deTeramond:2009xk}  - the positive contributions to $m^2_\pi$ from the LF potential and kinetic energy is cancelled by the constant term  in $U(\zeta^2,J)$ for $J=0.$   This holds for the positive sign of the dilaton profile  $\varphi(z) =  \kappa^2 z^2$.
The  LF dynamics retains conformal invariance of the action despite the presence of a fundamental mass scale.  
The constant term in the  LF potential $U(\zeta^2,J)$ derived from LF Holography is essential;  the masslessness of the pion and the separate dependence on $J$ and $L$ are  consequences of the potential derived from the holographic LF duality with AdS for general  $J$ and $L$~\cite{Brodsky:2013kpr, deTeramond:2013it}. Thus the light-front holographic approach provides an analytic frame-independent first approximation to the color-confining dynamics,  spectroscopy, and excitation spectra of the relativistic light-quark bound states of QCD.   It is systematically improvable in full QCD using  the basis light-front quantization (BLFQ) method~\cite{Vary:2009gt} and other methods. 

\begin{figure}[h]
\centering
\includegraphics[width=7.00cm]{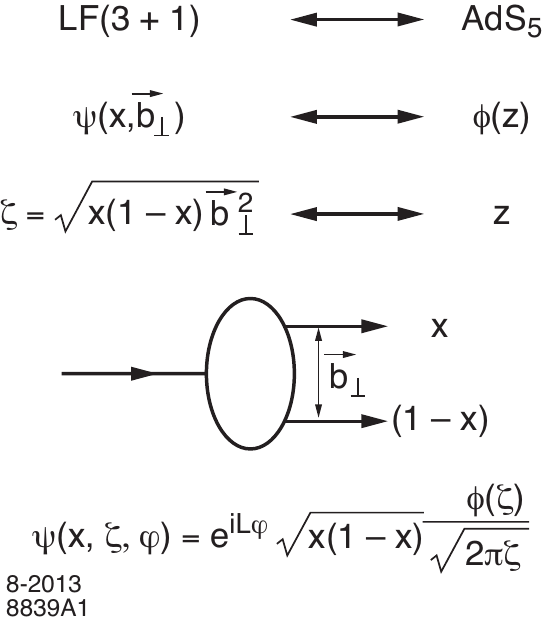}\hspace{0pt}
\caption{Light-Front Holography: Mapping between the hadronic wavefunctions of the Anti-de Sitter approach and eigensolutions of the light-front Hamiltonian theory  derived from the equality of LF and AdS  formula for EM and gravitational current matrix elements and their identical equations of motion.}
\label{dictionary}
\end{figure} 


We can now give an example of  light-front holographic mapping for the specific case of the elastic pion form factor.
In the higher-dimensional gravity theory, the hadronic transition amplitude  corresponds to
the  coupling of an external electromagnetic field $A^M(x,z)$,  for a photon propagating in AdS space, with an extended field $\Phi_P(x,z)$ describing a meson in AdS is~\cite{Polchinski:2002jw}
 \begin{equation} \label{MFF}
 \int \! d^4x \, dz  \sqrt{g} \, A^M(x,z)
 \Phi^*_{P'}(x,z) \overleftrightarrow\partial_M \Phi_P(x,z) \\
  \sim
 (2 \pi)^4 \delta^4 \left( P'  \! - P - q\right) \epsilon_\mu  (P + P')^\mu F_M(q^2) ,
 \end{equation}
where the coordinates of AdS$_5$ are the Minkowski coordinates $x^\mu$ and $z$ labeled $x^M = (x^\mu, z)$,
 with $M, N = 1, \cdots 5$,  and $g$ is the determinant of the metric tensor. 
The expression on the right-hand side
of (\ref{MFF}) represents the space-like QCD electromagnetic transition amplitude in physical space-time
$
\langle P' \vert J^\mu(0) \vert P \rangle = \left(P + P' \right)^\mu F_M(q^2).
$
It is the EM matrix element of the quark current  $J^\mu = \sum_q e_q \bar q \gamma^\mu q$, and represents a local coupling to pointlike constituents. Although the expressions for the transition amplitudes look very different, one can show  that a precise mapping of the matrix elements  can be carried out at fixed light-front time~\cite{Brodsky:2006uqa, Brodsky:2007hb}.

The form factor is computed in the light front formalism from the matrix elements of the plus current $J^+$
 in order to avoid coupling to Fock states with different numbers of constituents and is given by the Drell-Yan-West expression. The form factor can  be conveniently written in impact space as a sum of overlap of LFWFs of the $j = 1,2, \cdots, n-1$ spectator constituents~\cite{Soper:1976jc} 
\begin{equation} \label{eq:FFb}
F_M(q^2) =  \sum_n  \prod_{j=1}^{n-1}\int d x_j d^2 \mbf{b}_{\perp j}
\exp \! {\Bigl(i \mbf{q}_\perp \! \cdot \sum_{j=1}^{n-1} x_j \mbf{b}_{\perp j}\Bigr)}
\left\vert  \psi_{n/M}(x_j, \mbf{b}_{\perp j})\right\vert^2,
\end{equation}
corresponding to a change of transverse momentum $x_j \mbf{q}_\perp$ for each
of the $n-1$ spectators with  $\sum_{i = 1}^n \mbf{b}_{\perp i} = 0$.  The formula is exact if the sum is over all Fock states $n$.

For simplicity consider a two-parton bound-state. The $q \bar q$ LF Fock state wavefunction for a meson  can be written as
\begin{equation} \label{psi}
\psi(x,\zeta, \varphi) = e^{i L \varphi} X(x) \frac{\phi(\zeta)}{\sqrt{2 \pi \zeta}},
\end{equation}
thus factoring the longitudinal, $X(x)$,  transverse  $\phi(\zeta)$ and angular dependence $\varphi$.
 If both expressions for the form factor are to be
identical for arbitrary values of $Q$, we obtain $\phi(\zeta) = (\zeta/R)^{3/2} \Phi(\zeta)$ and $X(x) = \sqrt{x(1-x)}$~\cite{Brodsky:2006uqa},
where we identify the transverse impact LF variable $\zeta$ with the holographic variable $z$,
$z \to \zeta = \sqrt{x(1-x)} \vert \mbf b_\perp \vert$, where $x$ is the longitudinal momentum fraction and $ b_\perp$ is  the transverse-impact distance between the quark and antiquark. Extension of the results to arbitrary $n$ follows from the $x$-weighted definition of the transverse impact variable of the $n-1$ spectator system given in Ref.  \cite{Brodsky:2006uqa}.  Identical results follow from mapping the matrix elements of the energy-momentum tensor~\cite{Brodsky:2008pf}.

\section{The Light-Front Schr\"odinger Equation: A Semiclassical Approximation to QCD \label{LFQCD}}

It is advantageous to reduce the full multiparticle eigenvalue problem of the LF Hamiltonian to an effective light-front Schr\"odinger equation  which acts on the valence sector LF wavefunction and determines each eigensolution separately~\cite{Pauli:1998tf}.   In contrast,  diagonalizing the LF Hamiltonian yields all eigensolutions simultaneously, a complex task.
The central problem 
then becomes the derivation of the effective interaction $U$ which acts only on the valence sector of the theory and has, by definition, the same eigenvalue spectrum as the initial Hamiltonian problem.  In order to carry out this program one must systematically express the higher Fock components as functionals of the lower ones. This  method has the advantage that the Fock space is not truncated, and the symmetries of the Lagrangian are preserved~\cite{Pauli:1998tf}.

A hadron has four-momentum $P = (P^-, P^+,  \mbf{P}_\perp)$, $P^\pm = P^0 \pm P^3$ and invariant mass $P^2 = M^2$. The generators $P = (P^-, P^+,  \vec{P}_\perp)$ are constructed canonically from the QCD Lagrangian by quantizing the system on the light-front at fixed LF time $x^+$, $x^\pm = x^0 \pm x^3$~\cite{Brodsky:1997de}. 
The LF Hamiltonian $P^-$ generates the LF time evolution with respect to the LF time  $x^+$, whereas the LF longitudinal $P^+$ and transverse momentum $\vec P_\perp$ are kinematical generators.

In the limit of zero quark masses the longitudinal modes decouple  from the  invariant  LF Hamiltonian  equation  $H_{LF} \vert \phi \rangle  =  M^2 \vert \phi \rangle$,  with  $H_{LF} = P_\mu P^\mu  =  P^- P^+ -  \mbf{P}_\perp^2$.  The result is a relativistic and frame-independent light-front  wave equation for $\phi$~\cite{deTeramond:2008ht}
\begin{equation} \label{LFWE}
\left[-\frac{d^2}{d\zeta^2}
- \frac{1 - 4L^2}{4\zeta^2} + U\left(\zeta^2, J\right) \right]
\phi_{n,J,L}(\zeta^2) = 
M^2 \phi_{n,J,L}(\zeta^2).
\end{equation}
See fig. \ref{reduction}. This equation describes the spectrum of mesons as a function of $n$, the number of nodes in $\zeta$, the total angular momentum  $J$, which represent the maximum value of $\vert J^z \vert$, $J = \max \vert J^z \vert$,
and the internal orbital angular momentum of the constituents $L= \max \vert L^z\vert$.
The variable $z$ of AdS space is identified with the LF   boost-invariant transverse-impact variable $\zeta$~\cite{Brodsky:2006uqa}, 
thus giving the holographic variable a precise definition in LF QCD~\cite{deTeramond:2008ht, Brodsky:2006uqa}.
For a two-parton bound state $\zeta^2 = x(1-x) b^{\,2}_\perp$.
In the exact QCD theory $U$ is related to the two-particle irreducible $q \bar q$ Green's function.

\begin{figure}[h]
\centering
\includegraphics[width=15cm]{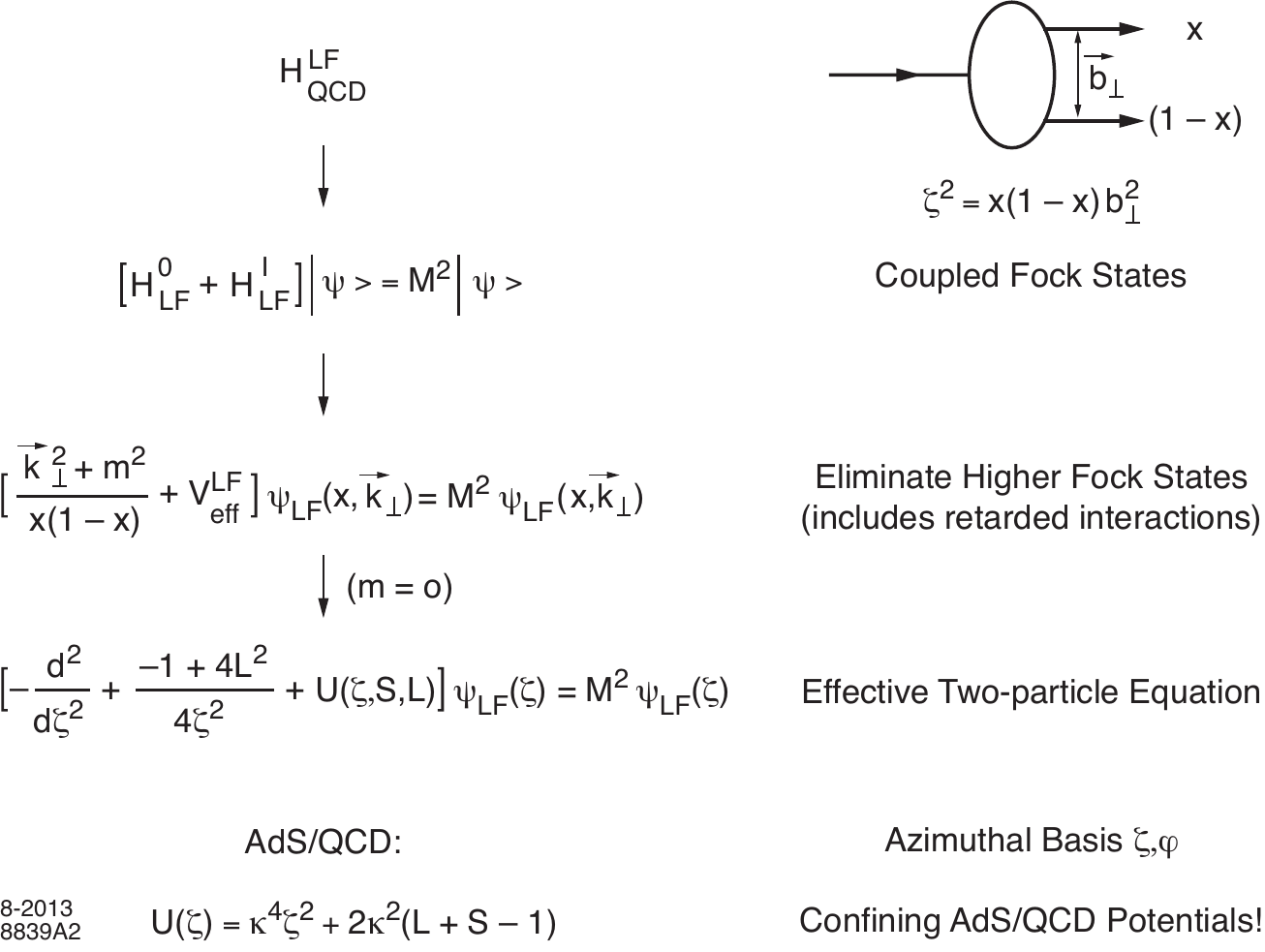}\hspace{0pt}
 \caption{Reduction of the QCD light-front Hamiltonian to an effective $q \bar q$ bound state equation. The potential is determined from spin-$J$ representations on AdS$_5$ space. The harmonic oscillator form of $U(\zeta^2)$ is determined by the requirement that the action remain conformally invariant.}
\label{reduction}
\end{figure} 

The potential in the the ``Light-Front Schr\"odinger equation (LFSE)'' is determined from the two-particle irreducible (2PI) $ q \bar q \to q \bar q $ Greens' function.  In particular, the reduction from higher Fock states in the intermediate states
leads to an effective interaction $U\left(\zeta^2, J\right)$  for the valence $\vert q \bar q \rangle$ Fock state~\cite{Pauli:1998tf}.
A related approach for determining the valence light-front wavefunction and studying the effects of higher Fock states without truncation has been given in Ref.~\cite{Chabysheva:2011ed}.

Unlike ordinary instant-time quantization, the light-front Hamiltonian equations of motion are frame independent; remarkably, they  have a structure which matches exactly the eigenmode equations in AdS space. This makes a direct connection of QCD with AdS methods possible.  In fact, one can
derive the light-front holographic duality of AdS  by starting from the light-front Hamiltonian equations of motion for a relativistic bound-state system
in physical space-time~\cite{deTeramond:2008ht}.

\section{Effective Confinement from the Gauge/Gravity Correspondence}

Recently we have derived wave equations for hadrons with arbitrary spin $J$ starting from an effective action in  AdS space~\cite{deTeramond:2013it}.    An essential element is the mapping of the higher-dimensional equations  to the LF Hamiltonian equation  found in Ref.~\cite {deTeramond:2008ht}.  This procedure allows a clear distinction between the kinematical and dynamical aspects of the LF holographic approach to hadron physics.  Accordingly, the non-trivial geometry of pure AdS space encodes the kinematics,  and the additional deformations of AdS encode the dynamics, including confinement~\cite{deTeramond:2013it}.

A spin-$J$ field in AdS$_{d+1}$ is represented by a rank $J$ tensor field $\Phi_{M_1 \cdots M_J}$, which is totally symmetric in all its indices.  In presence of a dilaton background field $\varphi(z)$ the effective action is~\cite{deTeramond:2013it} 
\begin{multline}
\label{Seff}
S_{\it eff} = \int d^{d} x \,dz \,\sqrt{\vert g \vert}  \; e^{\varphi(z)} \,g^{N_1 N_1'} \cdots  g^{N_J N_J'} \\ 
  \Big(  g^{M M'} D_M \Phi^*_{N_1 \dots N_J}\, D_{M'} \Phi_{N_1 ' \dots N_J'}  
 - \mu_{\it eff}^2(z)  \, \Phi^*_{N_1 \dots N_J} \, \Phi_{N_1 ' \dots N_J'} \Big),
 \end{multline}
where 
$D_M$ is the covariant derivative which includes parallel transport. 
The effective mass  $\mu_{\it eff}(z)$, which encodes kinematical aspects of the problem, is an {\it a priori} unknown function,  but the additional symmetry breaking due to its $z$-dependence allows a clear separation of kinematical and dynamical effects~\cite{deTeramond:2013it}.
The dilaton background field $\varphi(z)$ in  (\ref{Seff})   introduces an energy scale in the five-dimensional AdS action, thus breaking conformal invariance. It  vanishes in the conformal ultraviolet limit $z \to 0$.

 A physical hadron has plane-wave solutions and polarization indices along the 3 + 1 physical coordinates
 $\Phi_P(x,z)_{\nu_1 \cdots \nu_J} = e^{ i P \cdot x} \Phi_J(z) \epsilon_{\nu_1 \cdots \nu_J}({P})$,
 with four-momentum $P_\mu$ and  invariant hadronic mass  $P_\mu P^\mu \! = M^2$. All other components vanish identically. 
 The wave equations for hadronic modes follow from the Euler-Lagrange equation for tensors orthogonal to the holographic coordinate $z$,  $\Phi_{z N_2 \cdots N_J}  = 0$. Terms in the action which are linear in tensor fields, with one or more indices along the holographic direction, $\Phi_{z N_2 \cdots N_J}$, give us 
 the kinematical constraints required to eliminate the lower-spin states~\cite{deTeramond:2013it}.  Upon variation with respect to $ \hat \Phi^*_{\nu_1 \dots \nu_J}$,
 we find the equation of motion~\cite{deTeramond:2013it}  
\begin{equation}  \label{PhiJM}
 \left[ 
   -  \frac{ z^{d-1- 2J}}{e^{\varphi(z)}}   \partial_z \left(\frac{e^{\varphi(z)}}{z^{d-1-2J}} \partial_z   \right) \\
  +  \frac{(m\,R )^2}{z^2}  \right]  \Phi_J = M^2 \Phi_J,
  \end{equation}
  with  $(m \, R)^2 =(\mu_{\it eff}(z) R)^2  - J z \, \varphi'(z) + J(d - J +1)$,
  which is  the result found in Refs.~\cite{deTeramond:2008ht, deTeramond:2012rt} by rescaling the wave equation for a scalar field.  Similar results were found
  in Ref.~\cite{Gutsche:2011vb}. Upon variation with respect to
$ \hat \Phi^*_{N_1 \cdots z  \cdots N_J}$  we find the kinematical constraints which  eliminate lower spin states from the symmetric field tensor~\cite{deTeramond:2013it}  
\begin{equation} \label{sub-spin}
 \eta^{\mu \nu } P_\mu \,\epsilon_{\nu \nu_2 \cdots \nu_J}({P})=0, \quad
\eta^{\mu \nu } \,\epsilon_{\mu \nu \nu_3  \cdots \nu_J}({P})=0.
 \end{equation}

Upon the substitution of the holographic variable $z$ by the LF invariant variable $\zeta$ and replacing
  $\Phi_J(z)   = \left(R/z\right)^{J- (d-1)/2} e^{-\varphi(z)/2} \, \phi_J(z)$ 
in (\ref{PhiJM}), we find for $d=4$ the LF wave equation (\ref{LFWE}) with effective potential~\cite{deTeramond:2010ge}
\begin{equation} \label{U}
U(\zeta^2, J) = \frac{1}{2}\varphi''(\zeta^2) +\frac{1}{4} \varphi'(\zeta^2)^2  + \frac{2J - 3}{2 \zeta} \varphi'(\zeta^2) ,
\end{equation}
provided that the AdS mass $m$ in (\ref{PhiJM}) is related to the internal orbital angular momentum $L = max \vert L^z \vert$ and the total angular momentum $J^z = L^z + S^z$ according to $(m \, R)^2 = - (2-J)^2 + L^2$.  The critical value  $L=0$  corresponds to the lowest possible stable solution, the ground state of the LF Hamiltonian.
For $J = 0$ the five dimensional mass $m$
 is related to the orbital  momentum of the hadronic bound state by
 $(m \, R)^2 = - 4 + L^2$ and thus  $(m\, R)^2 \ge - 4$. The quantum mechanical stability condition $L^2 \ge 0$ is thus equivalent to the Breitenlohner-Freedman stability bound in AdS~\cite{Breitenlohner:1982jf}.

\begin{figure}[h]
\includegraphics[width=8.00cm]{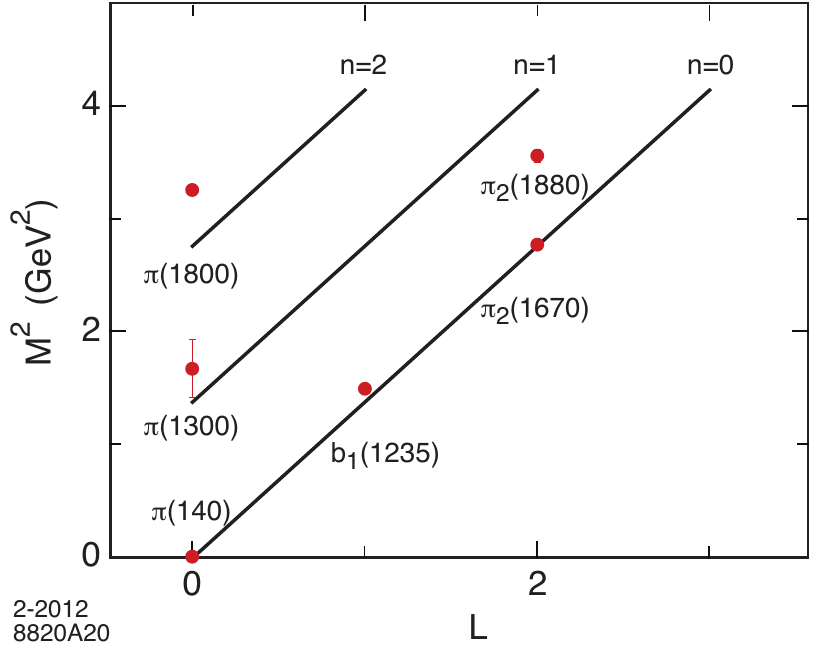} \hspace{-2pt}
\includegraphics[width=8.00cm]{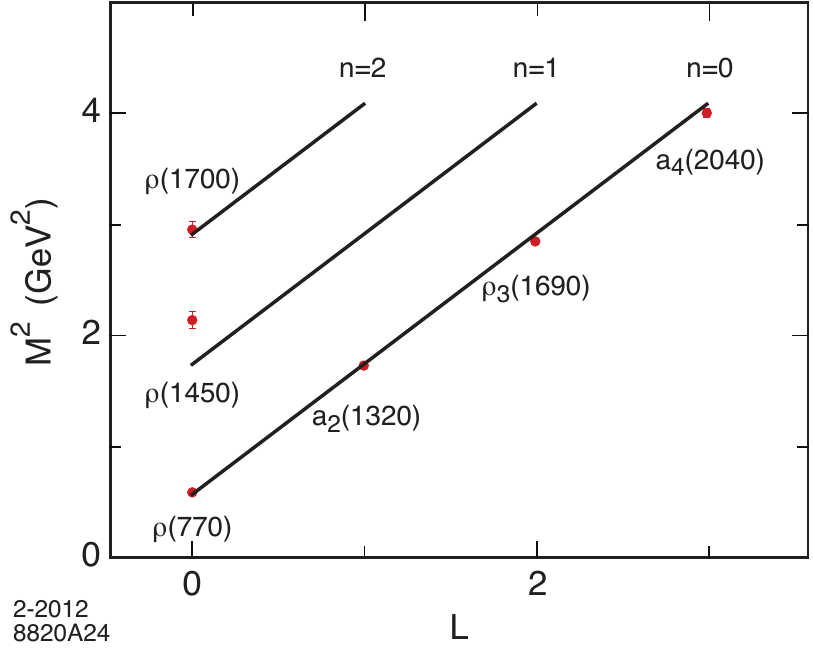}
 \caption{$I = 1$ parent and daughter Regge trajectories for the $\pi$-meson family (left) with
$\kappa= 0.59$ GeV; and  the   $\rho$-meson
 family (right) with $\kappa= 0.54$ GeV.}
\label{pionspec}
\end{figure} 

\begin{figure}[h]
\centering
\includegraphics[width=10.00cm]{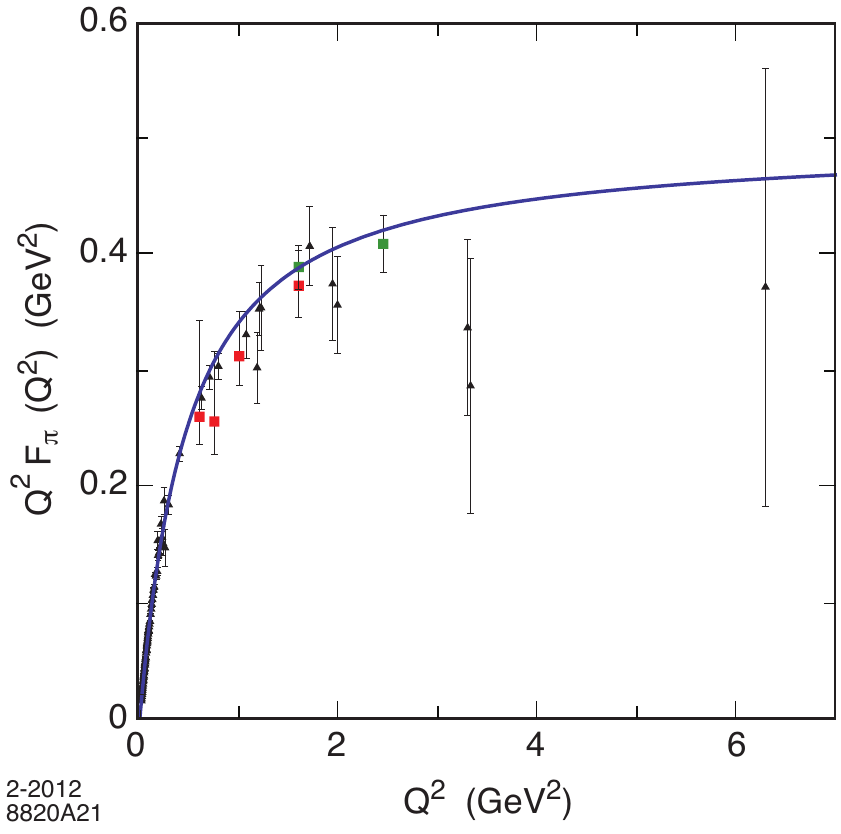} \hspace{0pt}
 \caption{Light-front holographic prediction for the space-like pion form factor.  }
\label{pionff}
\end{figure} 

The effective interaction $U(\zeta^2,J)$
is instantaneous in LF time and acts on the lowest state of the LF Hamiltonian.  This equation describes the spectrum of mesons as a function of $n$, the number of nodes in $\zeta^2$,
 the internal orbital angular momentum $L = L^z$, and the total angular momentum $J=J^z$,
with $J^z = L^z + S^z$  the sum of the  orbital angular momentum of the constituents and their internal spins.
The  ${\rm SO(2)}$ Casimir  $L^2$  corresponds to  the group of rotations in the transverse LF plane.
The LF wave equation is the relativistic frame-independent front-form analog of the non-relativistic radial Schr\"odinger equation for muonium  and other hydrogenic atoms in presence of an instantaneous Coulomb potential. The LF harmonic oscillator potential could in fact emerge from the exact QCD formulation when one includes contributions from the 
effective potential $U$ which are due to the exchange of two connected gluons; {\it i.e.}, ``H'' diagrams~\cite{Appelquist:1977tw}.
We notice that $U$ becomes complex for an excited state since a denominator can vanish; this gives a complex eigenvalue and the decay width.

The correspondence between the LF and AdS equations  thus determines the effective confining interaction $U$ in terms of the infrared behavior of AdS space and gives the holographic variable $z$ a kinematical interpretation. The identification of the orbital angular momentum 
is also a key element of our description of the internal structure of hadrons using holographic principles.

\begin{figure}[h]
\includegraphics[width=8.00cm]{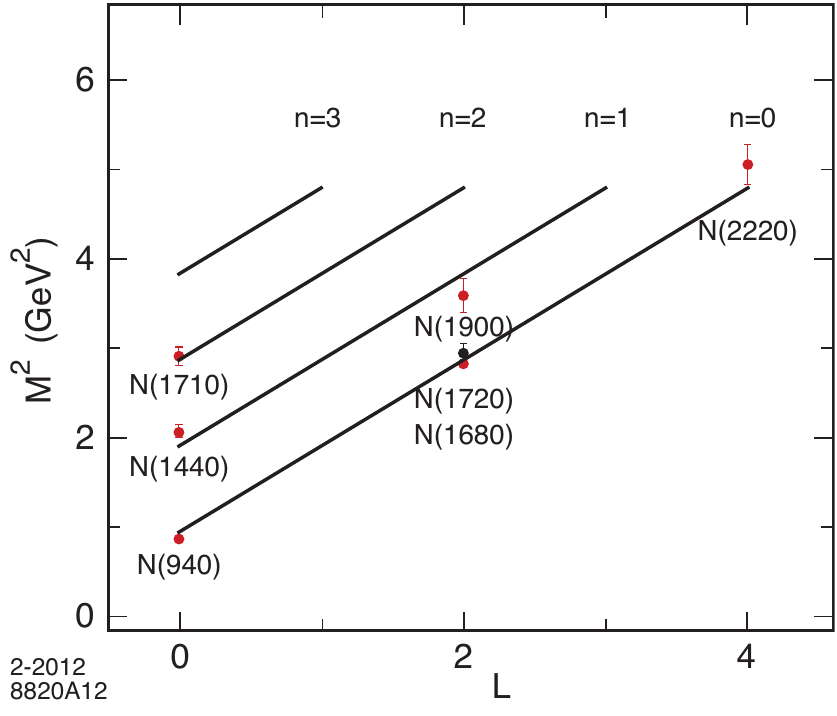} \hspace{-2pt}
\includegraphics[width=8.00cm]{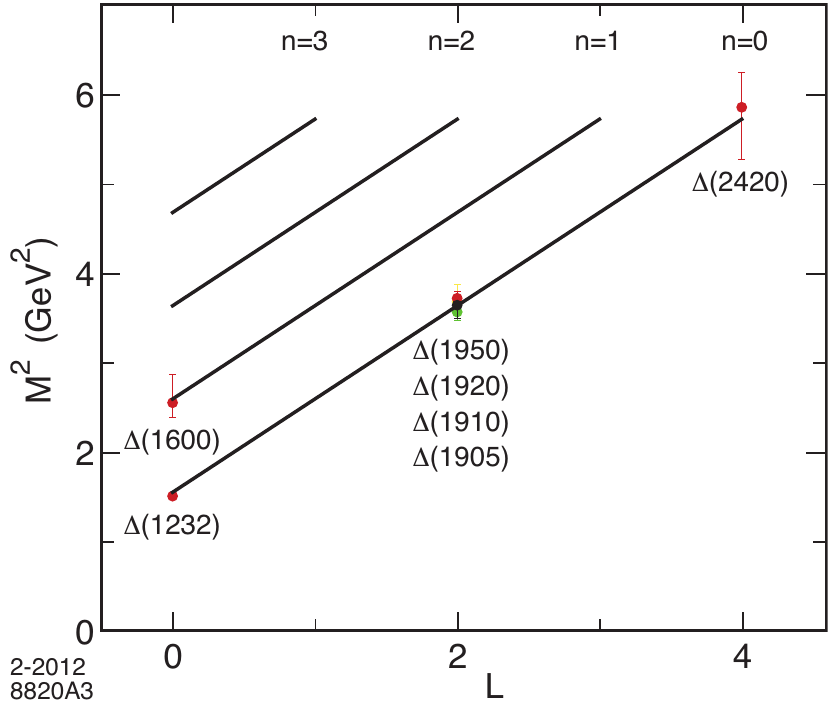} 
 \caption{Light front holographic predictions of the light-front Dirac equation for the nucleon spectrum. Orbital and radial excitations for the positive-parity sector are shown
 for the $N$ (left) and $\Delta$ (right) for $\kappa= 0.49$ GeV and $\kappa = 0.51$ GeV respectively. All confirmed positive and negative-parity resonances from PDG 2012 are well accounted using the procedure described in   \cite{deTeramond:2012rt}.}
\label{nucleonspect}
\end{figure} 

\begin{figure}[h]
\centering
\includegraphics[width=7.5cm]{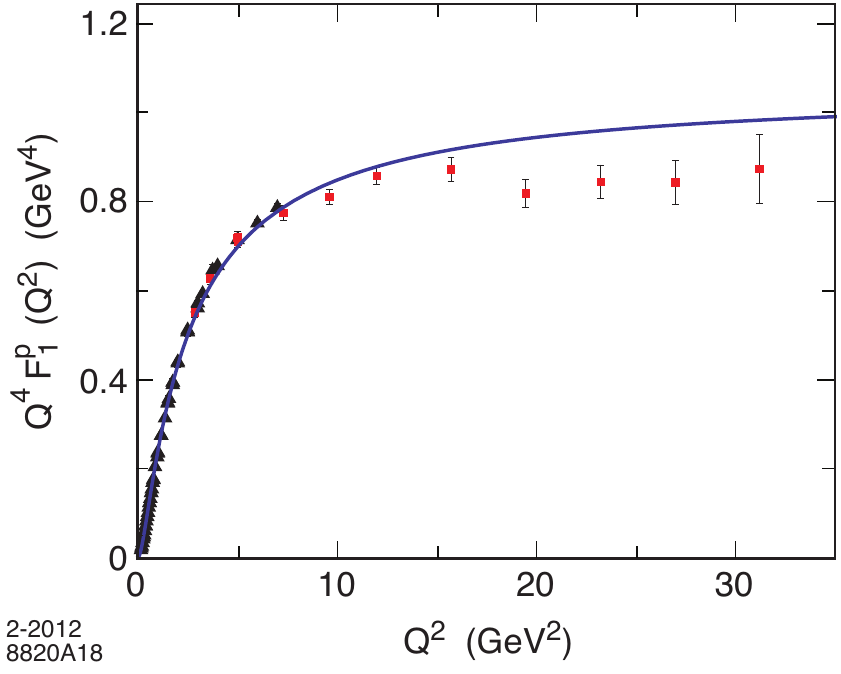}
 \hspace{-2pt}
\includegraphics[width=7.5cm]{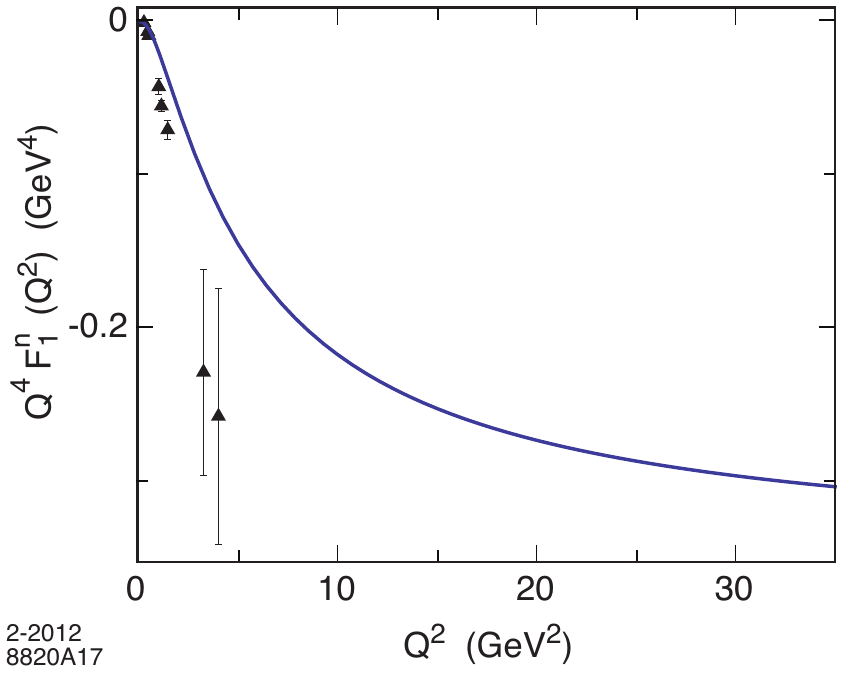}
\includegraphics[width=7.5cm]{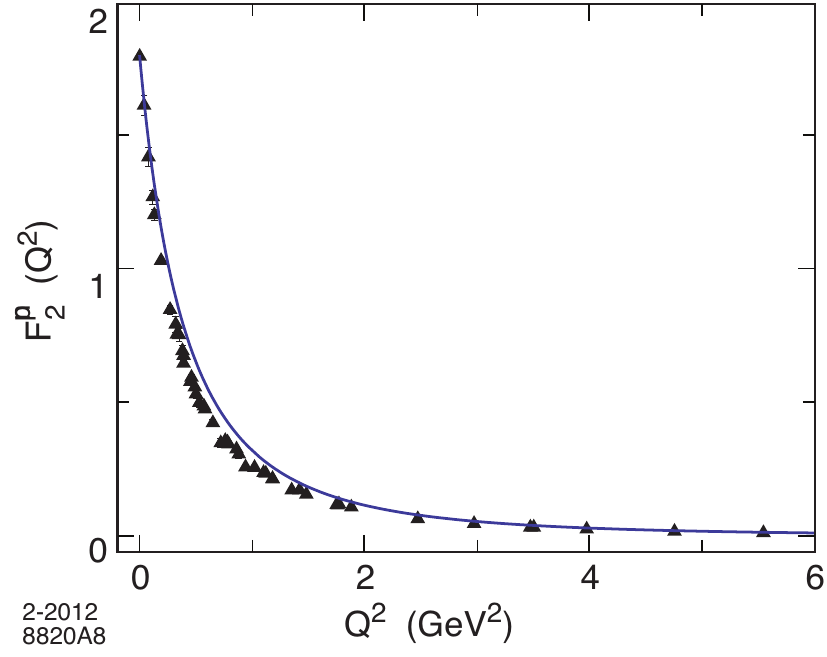} 
\hspace{-2pt}
\includegraphics[width=7.5cm]{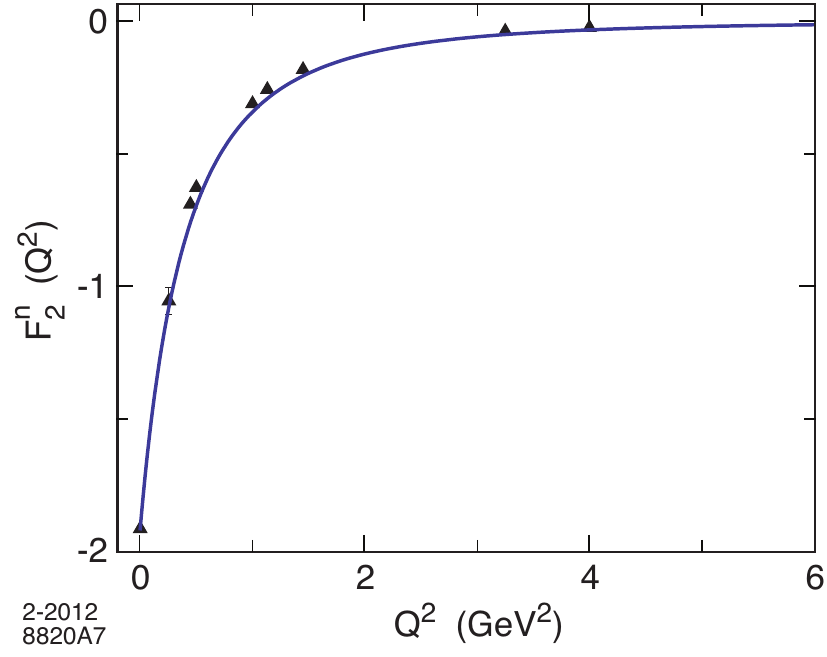} 
\caption{Light-front holographic predictions for the nucleon form factors normalized to their static values.  }
\label{nucleonffs}
\end{figure} 

The predictions of the resulting LF Schr\"odinger and Dirac equations for hadron light-quark spectroscopy and form factors for $m_q=0$ and $\kappa \simeq 0.5$ GeV are shown in Figs. \ref{pionspec}-\ref{nucleonffs} for a dilaton profile $\varphi(z) = \kappa^2 z^2$. A detailed discussion of the computations is given in Ref.  \cite{deTeramond:2012rt}.

\section{Uniqueness of the Confining Potential}

If one starts with a dilaton profile  $e^{\varphi(z)}$ with $\varphi \propto z^s,$  the existence of a massless pion in the limit of massless quarks determines
uniquely the value  $s = 2$.
To show this, one can use the stationarity of bound-state energies with respect to variation of parameters.
More generally, the effective theory should incorporate the fundamental conformal symmetry of the four-dimensional  classical QCD Lagrangian in the limit of massless quarks. To this end we study the invariance properties of a one-dimensional field theory under the full conformal group following the  dAFF construction of Hamiltonian operators described in Ref. ~\cite{deAlfaro:1976je}.

One starts with the one-dimensional action
$
{ \cal{S}}= \half \int dt (\dot Q^2 - g/Q^2),
$
which  is invariant under conformal transformations in the 
variable $t$. In addition to the Hamiltonian  $H_t$ there are two more invariants of motion for this field theory, namely the 
dilation operator $D$ and $K$, corresponding to the special conformal transformations in $t$.  
Specifically, if one introduces the  the new variable $\tau$ defined through 
$d\tau= d t/(u+v\,t + w\,t^2)$ and the  rescaled fields $q(\tau) = Q(t)/(u + v\, t + w \,t^2)^{1/2}$,
it then follows that the the operator
$G= u\,H_t + v\,D + w\,K$  generates
evolution in  $\tau$~\cite{deAlfaro:1976je}.
The Hamiltonian corresponding to the operator $G$ which introduces the mass scale
is a linear combination of the old Hamiltonian $H_t$, $D$, the generator of dilations, and
$K$, the generator of special conformal transformations.
It contains the confining potential 
$(4 u w - v^2) \zeta^2/8$, that is the confining term in (\ref{U}) for a quadratic dilaton profile and thus $\kappa^4 =  (4 u w - v^2)/8$.
The variable $\tau$ is related to the variable $t$ for the case  $u w >0, \,v=0$   by 
$
\tau =\frac{1}{\sqrt{u\,w}} \arctan\left(\sqrt{\frac{w}{u}} t\right),
$
{\it i.e.}, $\tau$ has only a limited range. The finite range of invariant LF time $\tau=x^+/P^+$ can be interpreted as a feature of the internal frame-independent LF  time difference between the confined constituents in a bound state. For example, in the collision of two mesons, it would allow one to compute the LF time difference between the two possible quark-quark collisions~\cite{Brodsky:2013kpr}.

\section{Summary}

The triple complementary connection of  (a)  AdS space,  (b) its LF holographic dual, 
and (c) the relation to the algebra of the conformal group in one dimension, is characterized by a quadratic confinement LF potential, and 
thus a dilaton profile with the power $z^s$, with the unique power $s = 2$. In fact,  for $s=2$ the mass of the $J=L=n=0$  pion is 
automatically zero in the chiral limit, and the separate dependence on $J$ and $L$ leads to a  mass ratio of the $\rho$ and the $a_1$ mesons which coincides with the result of the Weinberg sum rules~\cite{Weinberg:1967kj}. One predicts linear  Regge trajectories  with the same slope in the relative orbital angular momentum $L$ and the 
LF radial quantum humber $n$.  The AdS approach, however,  goes beyond the purely group theoretical considerations of dAFF, since 
features such as the masslessness of the pion and the separate dependence on $J$ and $L$ are a consequence of the potential (\ref{U}) derived from the duality with AdS for general high-spin representations.

\begin{figure}[h]
\label{alphas}
\centering
\includegraphics[width=10cm]{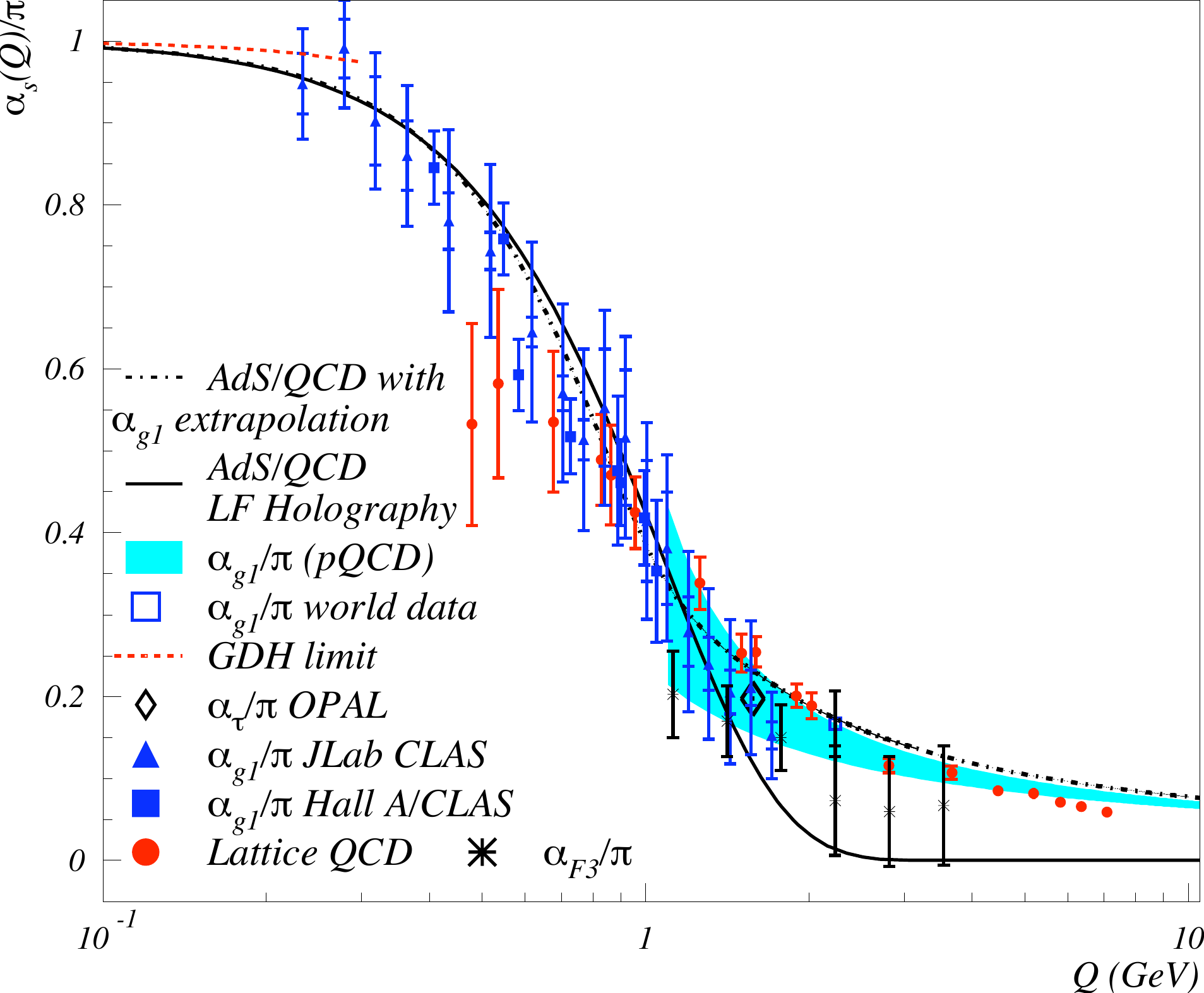} 
 \caption{Light-front holographic results for the QCD running coupling from Ref. ~\cite{Brodsky:2010ur} normalized to $\alpha_s(0)/ \pi =1.$  
 The result is analytic, defined at all scales and exhibits an infrared fixed point.}
 \label{alphas}
\end{figure} 

The QCD mass scale $\kappa$ in units of GeV has to be determined by one measurement; e.g., the pion decay constant $f_\pi.$  All other masses and size parameters are then predicted. The running of the QCD coupling  is predicted in the infrared region for $Q^2 <  4 \kappa^2$  to have the  form $\alpha_s(Q^2) \propto \exp{\left(-Q^2\over 4 \kappa^2\right)}$. As shown in Fig.  7, the result agrees with the shape of the effective charge defined from the Bjorken sum rule~\cite{Brodsky:2010ur}, displaying an infrared fixed point~\cite{Brodsky:2010ur}.  In the nonperturbative domain soft gluons are in effect sublimated into the effective confining potential. Above this region, hard-gluon exchange becomes important, leading to asymptotic freedom.  
The scheme-dependent scale $\Lambda_{QCD}$ that appears in the QCD running coupling in any given renormalization scheme  could be determined in terms of $\kappa.$

In our previous papers we have applied  LF holography to  baryon spectroscopy, space-like and time-like form factors, as well as transition amplitudes such as 
$\gamma^* \gamma \to \pi^0$, $\gamma^* N \to N^*$, all based on essentially one mass scale parameter $\kappa.$  Many other applications have been presented in the literature, including recent results by Forshaw and Sandapen~\cite{Forshaw:2012im} for diffractive  $\rho$ electroproduction based on the light-front holographic prediction for the longitudinal $\rho$ LFWF,  generalized parton distributions (GPDs)~\cite{Vega:2010ns}, and a model for nucleon and flavor form factors~\cite{Chakrabarti:2013dda}.

The treatment of the chiral limit in the LF holographic approach to strongly coupled QCD is substantially different from the standard approach  based on chiral perturbation theory.
In the conventional approach, 
spontaneous symmetry breaking  by a  non-vanishing chiral quark condensate $\langle  \bar \psi \psi  \rangle$ plays the crucial role.  In QCD sum  rules \cite{Shifman:1978bx}  $\langle  \bar \psi \psi \rangle$ brings in non-perturbative elements into the perturbatively calculated spectral sum rules. It should  be noted, however, that the definition of the condensate, even in lattice QCD necessitates a renormalization procedure for the operator product, and it is not a directly observable quantity.  In Bethe-Salpeter~\cite{Maris:1997hd} and light-front analyses~\cite{Brodsky:2012ku}, the Gell Mann-Oakes-Renner relation~\cite{GellMann:1968rz}
for $m^2_\pi/m_q$ involves the decay matrix element $\langle 0 |\bar \psi\gamma_5 \psi |\pi \rangle$ instead of $\langle 0| \bar \psi \psi | 0\rangle$.

In the color-confining 
light-front holographic model discussed here, the vanishing of the pion mass in the chiral limit, a phenomenon usually ascribed to spontaneous symmetry breaking of the chiral symmetry,  is  obtained specifically from the precise cancellation of the LF kinetic energy and LF potential energy terms for the quadratic confinement potential. This mechanism provides a  viable alternative to the conventional description of nonperturbative QCD based on vacuum condensates, and it 
eliminates a major conflict of hadron physics with the empirical value for the cosmological constant~\cite{Brodsky:2009zd,  Brodsky:2010xf}.

\section{Acknowledgments}
Invited talk, presented by SJB at Light-Cone 2012: Relativistic Hadronic and Particle Physics,
10 to 15 December, 2012 at the University of Delhi, New Delhi, India.
This work  was supported by the Department of Energy contract DE--AC02--76SF00515.

\end{document}